\documentclass[12pt]{article}
\usepackage{epsfig}
\begin{document}
\begin{titlepage}
\begin{center}

{\Large Nonextensive methods in turbulence and particle physics}

\vspace{2.cm}
{\bf Christian Beck}

\vspace{1cm}

School of Mathematical Sciences, Queen Mary,
University of London, Mile End Road, London E1 4NS, England

\end{center}

\vspace{3cm}

\abstract{We describe some recent applications of Tsallis
statistics in fully developed hydrodynamic turbulence and in high
energy physics. For many of these applications nonextensive
properties arise from spatial fluctuations of the temperature or
the energy dissipation rate. The entropic index $q$ is related to
the relative magnitude of these fluctuations. We concentrate on a
recently derived formula for the energy dependence of $q$ that is
experimentally verified by fits of cross sections in $e^+e^-$
annihilation experiments. Evaluating this formula for much smaller
energies $E$ of the order of the recombination temperature, one
obtains the correct order of magnitude of the fluctuations of the
cosmic microwave background. Evaluating it for $E\to \infty$, one
obtains (under certain assumptions) possible evidence for the
existence of 6 compactified dimensions, as predicted by
superstring theory.}

\end{titlepage}

\section{Introduction}

Nonextensive statistical mechanics as introduced by Tsallis in
1988  \cite{tsa1} and further developed by many others
\cite{abe}--\cite{tsa2} is
a useful generalization of ordinary statistical mechanics, with
potential applications not only for equilibrium systems but also
for nonequilibrium systems with a stationary state. It is based
on the extremization of the Tsallis entropies
\begin{equation}
S_q= \frac{1}{q-1} \left( 1- \sum_i p_i^q \right)
\end{equation}
rather than the Shannon entropy $S_1$, which is contained as a special
case in this more general setting. The $p_i$ are the probabilities
of the physical microstates, and $q$ is the entropic index.

Given two independent subsystems I and II with probabilities
$p_i^I$ and $p_i^{II}$, respectively, the Tsallis
entropy of the
composed system I+II (with probabilities
$p_{ij}^{I+II}=p_i^Ip_j^{II}$) satisfies
\begin{equation}
S_q^{I+ II}=S_q^{I}+S_q^{II}-(q-1)S_q^{I}S_q^{II} \label{nonad}.
\end{equation}
Hence the system is nonextensive, i.e. there is additivity of
the entropy for
$q=1$ only. 

Broadly speaking, the formalism with $q\not=1$ has so far been
observed to be relevant for three different classes of systems.
The first class is systems with long-range interactions
(e.g.~\cite{plastino}--\cite{latora}), the second one is
multifractal systems (e.g.~\cite{lyra}--\cite{lat2}), 
and the third one is systems
with fluctuations of temperature or energy dissipation
(e.g.~\cite{wilk}--\cite{benew1}).

In this paper we describe some interesting new physical
applications of the nonextensive approach. Our examples mainly
seem to be lying in the 3rd class. 
One example is fully developed
hydrodynamic turbulence \cite{beck01}--\cite{12}, another one the statistical
behaviour of particles produced in high energy $e^+e^-$ collider
experiments \cite{e+-1}--\cite{hage}. 
We will also deal with 
possible aspects of Tsallis statistics
in the early universe
\cite{rad1}--\cite{rad6}, and in particular point out a connection with
the fluctuations of the cosmic microwave background. In the last section
we will show that (under certain 
assumptions on the orign of the
temperature fluctuations) the $e^+e^-$ scattering
data may provide experimental evidence for the existence of
compactified dimensions, as predicted by superstring theory.

\section{Fluctuations of temperature and Tsallis sta\-tistics}

Let us here sketch the way in which fluctuations can lead to
nonextensive behaviour. This is a relatively new approach to
nonextensitivity \cite{wilk}--\cite{benew1}, which is in particular relevant for
nonequilibrium systems with a stationary state.
The fluctuations may be either local temperature fluctuations
or fluctuations of an effective friction coefficient (see
\cite{eddie1,eddie2} for some related nonequilibrium systems).

Consider a system of ordinary statistical mechanics with
Hamiltonian $H$. Tsallis statistics with $q>1$ can arise from this
ordinary Hamiltonian if the inverse temperature
$\beta$ is fluctuating. For example, we may think of these
fluctuations as spatial fluctuations.
%Using the integral represention of
%the gamma function
%\begin{equation}
%\Gamma (z)= \int_0^\infty e^{-t}t^{z-1}dt
%\end{equation}
%and substituting
%\begin{eqnarray}
%t&=&\beta\left( H+\frac{1}{(q-1)\beta_0}\right)\\ z&=&\frac{1}{q-1}
%\end{eqnarray}
From the integral representation of the gamma function one can easily
derive the formula \cite{wilk, beck01}
\begin{equation}
(1+(q-1)\beta_0 H)^{-\frac{1}{q-1}}= \int_0^\infty e^{-\beta
H} f(\beta ) d\beta \label{marl},
\end{equation}
where
\begin{equation}
f (\beta) = \frac{1}{\Gamma \left( \frac{1}{q-1} \right)} \left\{
\frac{1}{(q-1)\beta_0}\right\}^{\frac{1}{q-1}} \beta^{\frac{1}{q-1}-1}
\exp\left\{-\frac{\beta}{(q-1)\beta_0} \right\} \label{fluc}
\end{equation}
is the probability density of the $\chi^2$ (or gamma)
distribution. The above formula is valid for arbitrary
Hamiltonians $H$ und thus quite important. The left-hand side of
eq.~(\ref{marl}) is just the generalized Boltzmann factor emerging
out of nonextensive statistical mechanics. It can directly be
obtained by extremizing $S_q$. The right-hand side is a weighted
average over Boltzmann factors of ordinary statistical mechanics.
If we consider a nonequilibrim system with
fluctuating $\beta$ variables, then the generalized distribution
functions of nonextensive statistical mechanics are a consequence
of integrating over all possible fluctuating inverse temperatures
$\beta$, provided $\beta$ is $\chi^2$ distributed.

Wilk et al. \cite{wilk2} have also suggested an extension of this
approach 
which is valid for $q<1$. However, in that case
the distribution function $f(\beta)$ is more complicated and
different from the simple form (\ref{fluc}). In particular, for
$q<1$ one has $f(\beta)=f(\beta ,H)$, i.e.\ the probability
density of the fluctuating temperature also depends on the
Hamiltonian $H$. This case seems physically less straightforward,
hence we restrict ourselves to $q>1$ in the following.

For $q>1$, the $\chi^2$ distribution obtained in eq.~(\ref{fluc})
is a universal distribution that occurs in many very common
circumstances 
\cite{hast}. For example, it arises if $\beta$ is the sum of
squares of $n$ independent Gaussian random variables $X_j$
with average 0
\begin{equation}
\beta=\sum_{j=1}^n X_j^2, \label{nn}
\end{equation}
with
\begin{equation}
n=\frac{2}{q-1}. \label{e6}
\end{equation}
Hence one expects Tsallis statistics with $q>1$ to be
relevant in many applications where $n$ (approximately)
independent and (approximately) Gaussian variables contribute to
the variable $\beta$. Of course, other distribution functions
$f(\beta)$ can also be considered which may lead to other
generalized statistics. Tsallis statistics is just one possibility,
though a very relevant one. It is distinguished by generalized Khinchin
axioms and formal invariance of most thermodynamic relations
\cite{abe,3}.

%For fully developed turbulent hydrodynamic flows, where Tsallis
%statistics has been observed to work very well \cite{BLS},
%$\beta^{-1}$ is not the physical temperature of the flow but a
%formal temperature that is a function of the fluctuating energy
%dissipation rate in the flow times a suitable time scale
%\cite{beck01}. In the application to scattering processes in
%collider experiments \cite{e+-1,e+-2}, $\beta^{-1}$ is a
%fluctuating inverse temperature near the Hagedorn phase
%transition.

The constant $\beta_0$ in eq.~(\ref{fluc}) is the average of the
fluctuating $\beta$,
\begin{equation}
\langle \beta \rangle:=\int_0^\infty \beta f(\beta) d\beta =\beta_0
\end{equation}
The
deviation of $q$ from 1 can be related to the relative variance
of $\beta$. One can easily check that
\begin{equation}
q-1= \frac{\langle \beta^2 \rangle -\langle \beta\rangle^2}{\langle
\beta\rangle^2}.\label{q-1}
\end{equation}

\section{Application to turbulent flows}

A lot of recent papers \cite{beck01}--\cite{12}
showed that extremizing Tsallis entropies
is a very useful tool in fully developed turbulence,
leading to predictions for scaling exponents and probability
density functions that coincide quite precisely with experimental
data. Different models were developed, some of them using
Tsallis statistics with a constant $q<1$
within the multifractal model of turbulence \cite{ari1}--\cite{ari2},
others doing Tsallis statistics with
a scale-dependent $q>1$ directly in velocity space
\cite{hydro}--\cite{12}. Apparently,
the way Tsallis statistics can be used in turbulence is not
unique, but often the predictions given are quite similar.

Here, let us follow the approach introduced in \cite{hydro}
and experimentally verified in \cite{BLS}. Consider
turbulent velocity differences $u:=v(x+r)-v(x)$ at a certain scale
$r$ of the flow. Consider a nonlinear Langevin equation
of the form
\begin{equation}
\dot{u}=-\gamma F(u)+\sigma L(t) \label{there}
\end{equation}
where $F(u)=-\frac{\partial}{\partial u}V(u)$ is a nonlinear
forcing and $L(t)$ is Gaussian white noise. To be specific, let us
assume that $V(u)=C|u|^{2\alpha}$ is a power-law potential. The
parameter $\beta$ is defined as
\begin{equation}
\beta=\frac{\gamma}{\sigma^2}.
\end{equation}
For $\alpha =1$ we just obtain the ordinary Ornstein-Uhlenbeck
process, for which $\beta=\gamma/\sigma^2$
is related to the temperature of the surrounding heat bath
by $\beta=1/kT$. Given some $\beta$
and $\alpha$, the stochastic differential equation (\ref{there})
generates the stationary probability density
\begin{equation}
p(u|\beta)=\frac{\alpha}{\Gamma \left( \frac{1}{2\alpha}\right)}
\left( C\beta \right)^\frac{1}{2\alpha}\exp\left\{- \beta C
|u|^{2\alpha}\right\} .
\end{equation}
So far this
invariant density deviates from what is measured in turbulence
experiments, so our model is yet too simple for turbulence.
But now comes the important point. Let us assume that
not only $L(t)$ is a fluctuating quantity but also
$\beta=\gamma /\sigma^2$. We
assume that $\beta$ changes spatially on a relatively large scale,
i.e. the test particle first moves in a certain region with a
given $\beta$, then moves to another region with another $\beta$
and so on. Within each region
there is local equilibrium.
If $\beta$ is a $\chi^2$ distributed random variable in the various regions,
with an $n$ given by eq.~(\ref{nn}),
then for the marginal distributions $p(u)=\int
p(u|\beta)f(\beta)d\beta$ we obtain after a short calculation
\begin{equation}
p(u)=\frac{1}{Z_q}\frac{1}{(1+(q-1)\tilde{\beta}C
|u|^{2\alpha})^{\frac{1}{q-1}}} \label{pu},
\end{equation}
where
\begin{equation}
Z_q^{-1}= \alpha \left( C(q-1)\tilde{\beta}\right)^{\frac{1}{2\alpha}}
\cdot \frac{\Gamma \left( \frac{1}{q-1}\right)}{ \Gamma
\left( \frac{1}{2\alpha} \right) \Gamma \left(
\frac{1}{q-1}-\frac{1}{2\alpha} \right)}
\end{equation}
and
\begin{eqnarray}
q&=&1+\frac{2\alpha}{\alpha n+1} \label{qn} \\
\tilde{\beta}&=&\frac{2\alpha}{1+2\alpha -q}\beta_0.
\end{eqnarray}
The densities (\ref{pu}) extremize the Tsallis entropies.
Hence Tsallis statistics is rigorously proved for this simple
model with fluctuating friction forces
\cite{benew1}. Moreover, for a test particle 
moving in a turbulent flow
the dynamics is physically plausible, since one knows that the dissipation
fluctuates in turbulent flows.

The
densities $p(u)$ as given by eq.~(\ref{pu}) were just those that
were recently successfully
used as fits of experimentally measured turbulent densities 
\cite{BLS}. They perfectly coincide with the
experimental data (Fig.~1 in \cite{BLS},
Fig.~1 in \cite{beck01}), thus making this nonextensive model
a very good model of turbulent
statistics. The generalized
Langevin equation yields a dynamical reason
for Tsallis statistics. If one replaces the Gaussian white noise in the above
stochastic differential equation by deterministic chaotic noise,
one can also understand the (very small) asymmetry (skewness) of
the observed distributions \cite{hilgers, hi2, hydro, BLS}.

\section{Application to $e^+e^-$ annihilation}

Not only velocity distributions in hydrodynamics,
but also momentum distributions of particles produced
in high-energy scattering experiments can be successfully
described using nonextensive methods.
The
differential cross section $\frac{1}{\sigma}
\frac{d\sigma}{dp}$ measured in a collider experiment is proportional
to the probability density to observe particles
with a certain momentum $p$. Bediaga
et al. \cite{e+-1} looked at experimentally measured differential cross
sections of transverse momenta $p_T$ in $e^+e^-$ annihilation experiments
and
showed that the formula
\begin{equation}
\frac{1}{\sigma} \frac{d\sigma}{dp_T} = c \cdot u \int_0^\infty dx \;
\left( 1+(q-1)\sqrt{x^2+u^2+m_\beta^2} \right)^{-\frac{q}{q-1}}
\label{exact} \end{equation} yields an extremely good fit of the
experimentally measured cross sections. Here $x=p_L/kT_0$, $u=p_T/kT_0$ and
$m_\beta:=m_0/kT_0$ are the longitudinal momentum, transverse
momentum and mass of the particles in units of the Hagedorn temperature $T_0$,
respectively.
$c$ is a
constant related to the multiplicity $M$, i.e.\ the number of
charged particles produced.

The Hagedorn temperature $T_0$ is a kind of phase
transition temperature of `boiling' nuclear matter. This
fundamental concept
was introduced by Hagedorn in a seminal paper in
1965 \cite{hage,ion1,seek}. These
days the Hagedorn phase transition is seen in connection
with the QCD phase transition
from confined to non-confined states.

Bediaga et al. just wrote down eq.~(\ref{exact}) without deriving
it. They neglected $m_\beta$
and used three free fitting parameters for their fits, namely 
$c, T_0, q$.
The question was if one can derive suitable equations
for these fitting parameters, thus proceeding from a mere fit to a theory.
Although the complete theory taking into
account all possible interactions within the
hadronization cascade is out of reach, a first step
was performed in \cite{e+-2}, where eq.~(\ref{exact}) was derived
from first principles
by $q$-generalizing Hagedorn's original approach.
The tool for this are
suitable factorizing nonextensive partition functions for fermions and
bosons, actually of similar form as previously suggested in 
\cite{rad3}.

One can then go a step further
by considering suitable approximations and by
making plausible assumptions on the energy dependences of 
various parameters, being guided by the existence of moments.
The final result of \cite{e+-2} was that in good approximation one has
\begin{equation}
\frac{1}{\sigma}\frac{d\sigma}{dp_T}= \frac{1}{T_0} M \cdot p(u)
\label{here7}
\end{equation}
where
\begin{eqnarray}
p(u)&=& \frac{1}{Z_q}
u^{3/2} \left( 1+(q-1)u \right)^{-\frac{q}{q-1}+\frac{1}{2}}
\label{here3} \\
Z_q&=&
(q-1)^{-5/2} B\left( \frac{5}{2},
\frac{q}{q-1}-3 \right) \\
T_0(q)&=&\left( 1-\frac{q}{3} \right) p_T^* \label{tqq}\\
p_T^* &\approx& 180 \; MeV \\
q(E)&=&\frac{11-e^{-E/E_0}}{9+e^{-E/E_0}}  \label{here10} \\
E_0 &\approx &\frac{1}{2}m_Z =45.6 \; GeV  \label{e23} \\
M(E) &\approx&\left( \frac{E}{T_0(1)}\right)^{5/11} .
\label{here88}
\end{eqnarray}
Here $E$ is the center of mass energy of the beam, and
the function $B(x,y)=\Gamma(x)\Gamma(y)/\Gamma(x+y)$ denotes
the beta function.

Formula (\ref{here7}) with $p(u)$ given by
eq.~(\ref{here3}), $q(E)$ given by eq.~(\ref{here10}),
$T_0(q)$ given by eq.~(\ref{tqq})
and multiplicity $M(E)$ given by eq.~(\ref{here88})
turns out to very well reproduce
the experimental results of cross sections
for all energies $E$ in the experimentally
accessible region 14...161 GeV (see
Fig.~3 in \cite{e+-2}, which compares with experimental cross sections
measured by the
TASSO and DELPHI collaborations \cite{TASSO, DELPHI}).

The theoretical progress is
that eqs.~(\ref{here7})--(\ref{here88}) provide
concrete equations for the energy dependence of all parameters
$T_0,M,q$ used in the fits, and only two fundamental constants are left:
One is $p_T^*$, the energy scale of the QCD phase transition,
and the other one $E_0$, the energy scale of the electroweak
phase transition.

The agreement of eq.~(\ref{here7})
with the experimentally measured
cross sections is remarkably good. Nonextensitivity
seems to be the right approach to tackle this high energy problem.
$q$ is not constant but depends on the spatial scale $r=\hbar/E$,
just similar as it does in turbulence.
Of course, an alternative for particle physicists is to do Monte-Carlo
simulations. The simulated Monte-Carlo cross sections
look very similar to our theoretical curves 
\cite{passon}.
Thus the complexity inherent in the Monte-Carlo
simulation effectively reduces to some
model with nonextensive properties.

\section{Fluctuations of the cosmic microwave background}

Many authors have already dealt with 
possible applications of nonextensive statistical mechanics in
high energy physics. Work so far has concentrated
on the solar neutrino problem \cite{lav0}--\cite{sta2}, quark
gluon plasmas
\cite{raf}, heavy ion collisions \cite{qua}, cosmic rays \cite{wilk3},
$e^+e^-$ annihilation \cite{e+-1,e+-2}, 
and the cosmic microwave background \cite{rad1,rad3}.
Useful bounds on $|q-1|$ in the early universe were
derived \cite{rad1}--\cite{rad6}. 
Here we want to deal with the cosmic microwave background 
from a new point of view.
We will estimate the order of magnitude of
the temperature fluctuations using our formula for $q(E)$ from the previous
section.

Although formula (\ref{here10}) is only experimentally confirmed
for energies in the region 14...161 GeV, we can certainly also
evaluate it for other energies as well. For $E<<E_0$ we may write
$exp(-E/E_0)\approx 1-E/E_0$, which yields
\begin{equation}
q(E)=1+\frac{1}{5}\frac{E}{E_0} \;\;\;\;\; (E<<E_0). \label{e25}
\end{equation}
In section 2 we saw that any $q>1$ can be interpreted
in terms of spatial fluctuations in $\beta
=1/kT$. It is now interesting to see what the typical size of
these fluctuations is if we evaluate
our formulas at the temperature $E\approx 1$ eV where
the universe becomes transparent (the
recombination temperature). One obtains from formulas
(\ref{q-1}), (\ref{e23}), (\ref{e25})
\begin{equation}
\frac{\Delta{\beta}}{\beta_0}=\frac{\Delta T}{T}=\sqrt{q-1}
=\sqrt{\frac{E}{5E_0}}\approx 2 \cdot 10^{-6}.
\end{equation}
Here $\Delta \beta$ denotes the standard deviation of $\beta$.
Remarkably, we obtain a value that has the same
order of magnitude as the experimentally measured 
value of the fluctuations of the cosmic
microwave background \cite{bennet}:
\begin{equation}
1.5 \cdot 10^{-6} \leq \frac{\Delta T}{T} \leq 10 \cdot 10^{-6}
\end{equation}
Of course,
the correct order of magnitude  
may just be a random coincidence, but it may also
not be!
If true, then nonextensive statistical mechanics
is probably relevant for the thermodynamic description of the
early universe, due to fluctuations that were neglected so far.
We expect that
$q$ is energy dependent
and given by eq.~(\ref{here10}) or something of similar order of
magnitude.
Our value of $q-1 \approx 4\cdot 10^{-12}$ at the recombination temperature
is certainly consistent with previous bounds derived, which
were of the order
$|q-1| \leq   10^{-4}$ \cite{rad1,rad2,rad4,rad6}. 

\section{Compactified dimensions}

Another striking observation is related to the $E\to \infty$
limit, where our formula (\ref{here10}) yields
\begin{equation}
q(\infty)=\frac{11}{9}.
\end{equation}
The largest possible energy $E$ corresponds to the smallest
possible scale $ r\sim \hbar/E$. In a quantum
gravity setting this is the Planck
scale. It is now reasonable to assume that at the smallest
possible scale $r$ (described by the largest possible energy $E$)
the number $n$ of {\em independent} contributions $X_j$ to the
fluctuating  $\beta$ is given by the number $d$ of spatial
dimensions of the space. The reason is that the momenta whose
kinetic energy define $\beta$ can fluctuate in each space
direction independently. At the smallest scales, there is (by
definition of the smallest scale) no other
structure and degree of freedom except for the various directions in 
$d$-dimensional space. Using
this (somewhat speculative) argument we obtain from eq.~(\ref{e6})
\begin{equation}
q-1= \frac{2}{n}=\frac{2}{d}=\frac{2}{9}
\;\;\;\;\;\;\;\;\mbox{(at smallest scales),} \label{dimqq}
\end{equation}
hence $d=9$ at the smallest scales. 9 spatial
dimensions mean that space-time is 10-dimensional. Remarkably, we
get the number of space-time dimensions that are necessary for
superstring theory to be formulated in a consistent way
\cite{GSW}.

The $e^+e^-$ annihilation data actually already show a crossover
from $q\approx 1$ to $q\approx 11/9$ at the electroweak energy
scale of about 100 GeV \cite{e+-1, e+-2}. Should the saturation of $q$ near the
value $11/9$ be confirmed by future experiments with larger center
of mass energies $E>161$ GeV, this could be interpreted as
possible experimental evidence that the compactified extra
dimensions of superstring
theory start already becoming visible at the electroweak scale.
This means, their diameter would be of the order of $r\sim \hbar c (100
GeV)^{-1}$, much larger than the Planck length. This is compatible
with other theoretical arguments, which also suggest relatively
large sizes of the extra dimensions \cite{anto}.
If this picture is correct,
then the ultimate reason for the validity of
Tsallis statistics in the early universe could be 
fluctuations originating 
from the Planck era, where $q=11/9$.


\begin{thebibliography}{99}
\bibitem{tsa1} C. Tsallis, J. Stat. Phys. {\bf 52}, 479 (1988)
\bibitem{abe} S. Abe, Phys. Lett. {\bf 271A}, 74 (2000)
\bibitem{3} C. Tsallis, Braz. J. Phys. {\bf 29}, 1 (1999)
\bibitem{cura} E.M.F. Curado and C. Tsallis, J. Phys. {\bf 24A}, L69 (1991)
\bibitem{plasti} A.R. Plastino and A. Plastino, Phys. Lett. {\bf 177A},
177 (1993)
\bibitem{abe2} S. Abe, S. Mart\'{i}nez, F. Pennini,
and A. Plastino, Phys. Lett. {\bf 281A}, 126 (2001)
\bibitem{abe3} S. Abe, Phys. Rev. {\bf 63E}, 061105 (2001)
\bibitem{tsa2} C. Tsallis,  R.S. Mendes and A.R. Plastino, Physica {\bf 261A},
534 (1998)
\bibitem{plastino} A.R. Plastino and A. Plastino,
Phys. Lett. {\bf 174A}, 384 (1993)
\bibitem{astro2} A. Lavagno, G. Kaniadakis, M. Rego-Monteiro,
P. Quarati and C. Tsallis, Astrophys. Lett. and Comm. {\bf 35}, 449 (1998)
\bibitem{land} P.T. Landsberg, J. Stat. Phys. {\bf 35}, 159 (1984)
\bibitem{ruffo} M. Antoni and S. Ruffo, Phys. Rev. {\bf 52E}, 2361 (1995)
\bibitem{latora} V. Latora, A. Rapisarda and C. Tsallis, cond-mat/0103540
\bibitem{lyra} M.L. Lyra and C. Tsallis, Phys. Rev. Lett. {\bf 80}, 53 (1998)
\bibitem{lat2} V. Latora, M. Baranger, A. Rapisarda, C. Tsallis,
Phys. Lett. {\bf 273A}, 97 (2000)
\bibitem{wilk} G. Wilk and Z. Wlodarczyk, Phys. Rev. Lett. {\bf 84}, 2770 (2000)
\bibitem{wilk2} G. Wilk and Z. Wlodarczyk, hep-ph/0004250,
to appear in Chaos, Solitons and Fractals (2001)
\bibitem{beck01} C. Beck, Phys. Lett. {\bf 287A}, 240 (2001)
\bibitem{benew2} C. Beck, to appear in Europhys. Lett. (2001)
(cond-mat/0105371)
\bibitem{benew1} C. Beck, to appear in Phys. Rev. Lett. (2001)
(cond-mat/0105374)
%\bibitem{web} http://tsallis.cat.cbpf.br/biblio.htm
\bibitem{ari1} T. Arimitsu and N. Arimitsu, J. Phys. {\bf 33A}, L235 (2000)
\bibitem{ari3} T. Arimitsu and N. Arimitsu, Prog. Theor. Phys. {\bf 105},
355 (2001)
\bibitem{ari2} T. Arimitsu and N. Arimitsu, Phys. Rev. {\bf 61E}, 3237 (2000)
\bibitem{hydro} C. Beck, Physica {\bf 277A}, 115 (2000)
\bibitem{BLS} C. Beck, G.S. Lewis and H.L. Swinney, Phys. Rev. {\bf 63E},
 035303(R) (2001)
\bibitem{becks} C. Beck, Physica {\bf 295A}, 195 (2001)
\bibitem{shiva} B.K. Shivamoggi, C. Beck, J. Phys. {\bf 34A},
4003 (2001)
\bibitem{12} F.M. Ramos, C. Rodrigues Neto and R.R. Rosa,
cond-mat/0010435
%\bibitem{vKa} N.G. van Kampen, {\em Stochastic Processes in Physics
%and Chemistry}, North-Holland, Amsterdam (1981)
%\bibitem{BS} C. Beck and F. Schl\"{o}gl, {\em Thermodynamics of
%Chaotic Systems}, Camdridge University Press, Cambridge (1993)
%\bibitem{swinney} G.S. Lewis and H.L. Swinney, Phys. Rev. {\bf 59E},
%5457 (1999)
\bibitem{e+-1} I. Bediaga, E.M.F. Curado, and J. Miranda,
Physica {\bf 286A}, 156 (2000)
\bibitem{e+-2} C. Beck, Physica {\bf 286A}, 164 (2000)
\bibitem{hage} R. Hagedorn, Suppl. Nuovo Cim. {\bf 3}, 147 (1965)
\bibitem{rad1} C. Tsallis, F.C. Sa Barreto, E.D. Loh,
Phys. Rev. {\bf 52E}, 1447 (1995)
\bibitem{rad2} A.R. Plastino, A. Plastino, H. Vucetich, Phys. Lett. {\bf 207A},
42 (1995)
\bibitem{rad3} U. Tirnakli, F. B\"uy\"ukkilic, D. Demirhan,
Physica {\bf 240A}, 657 (1997)
\bibitem{rad4} D.F. Torres, H. Vucetich, A. Plastino, Phys. Rev. Lett. {\bf 79},
1588 (1997) 
\bibitem{rad5} U. Tirnakli, D.F. Torres, Physica {\bf 268A}, 225 (1999)
\bibitem{rad6} U. Tirnakli, F. B\"uy\"ukkilic, D. Demirhan,
Phys. Lett. {\bf 245A}, 62 (1998)
\bibitem{eddie1} E.G.D. Cohen, Physica {\bf 240A}, 45 (1997)
\bibitem{eddie2} G. Gallavotti, E.G.D. Cohen, Phys. Rev. Lett. {\bf 74}, 2694
(1995)
\bibitem{hast} N.A.J. Hastings and J.B. Peacock,
{\em Statistical Distributions}, Butterworth, London (1974)
\bibitem{hilgers} A. Hilgers, C. Beck, Phys. Rev. {\bf 60E}, 5385 (1999)
\bibitem{hi2} A. Hilgers, C. Beck, Physica {\bf 156D}, 1 (2001)
\bibitem{ion1} R. Hagedorn and J. Rafelski, Phys. Lett. {\bf 97B}, 136 (1980)
\bibitem{seek} P.T. Landsberg, {\em Seeking Ultimates}, IOP, Bristol (2000)
\bibitem{TASSO} TASSO collaboration, Z. Phys. {\bf 22C}, 307 (1984)
\bibitem{DELPHI} DELPHI collaboration, Z. Phys. {\bf 73C}, 229 (1997)
\bibitem{passon} O. Passon, 
Diplomarbeit, Universit\"at Wuppertal (1997)
\bibitem{lav0} G. Kaniadakis, A. Lavagno, P. Quarati,
Phys. Lett. {\bf 369B}, 308 (1996)
\bibitem{lav} A. Lavagno, P. Quarati, Phys. Lett. {\bf 498B}, 47 (2001)
\bibitem{sta1} N.L. Aleksandrov, A.N. Starostin, J. Exp. Theor. Phys. {\bf 86}, 
903 (1998)
\bibitem{sta2} A.N. Starostin, V.I. Savchenko, N.J. Fisch, Phys. Lett. {\bf
274A}, 64 (2000)
\bibitem{raf} D.B. Walton and J. Rafelski, Phys. Rev. Lett. {\bf 84}, 31 (2000)
\bibitem{qua} W.M. Alberico, A. Lavagno, P. Quarati, Eur. Phys. J. {\bf C12},
499 (2000)
\bibitem{wilk3} G. Wilk and Z. Wlodarczyk, Nucl. Phys. B (Proc. Suppl.)
{\bf A75}, 191 (1999)
\bibitem{bennet} C.L. Bennett et al., Astrophys. J. {\bf 464}, L1 (1996)
\bibitem{GSW}  M.B. Green, J.H. Schwarz, E. Witten, {\em Superstring
Theory}, vol. I and II, Cambridge University Press, Cambridge (1987)
\bibitem{anto} I. Antoniadis, K. Benakliand, M. Quiros, Phys. Lett.
{\bf 460B}, 176 (1999)
\end{thebibliography}
\end{document}